\documentclass{sig-alternate-2013}
\usepackage[utf8]{inputenc}
\usepackage{algorithmic}
\usepackage{algorithm}
\usepackage{todonotes}
\usepackage{etoolbox, siunitx, booktabs}
\usepackage[moderate]{savetrees}
\robustify\bfseries
\usepackage{colortbl}
\usepackage{float}
\usepackage{multirow}
\newfloat{algorithm}{t}{lop}

\permission{\copyright 2017 International World Wide Web Conference Committee \\ (IW3C2), published under Creative Commons CC BY 4.0 License.}
\conferenceinfo{WWW 2017,}{April 3--7, 2017, Perth, Australia.}
\copyrightetc{ACM \the\acmcopyr}
\crdata{978-1-4503-4913-0/17/04. \\
http://dx.doi.org/10.1145/3041021.3054251    \\
\includegraphics{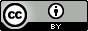}
}

\begin{document}






%

\title{Time-Series Adaptive Estimation of Vaccination Uptake Using Web Search Queries}
%
%
%
%
%

\numberofauthors{4} 
%
\author{
%
%
\alignauthor Niels Dalum Hansen\\
       \affaddr{University of Copenhagen/IBM}\\
       \email{nhansen@di.ku.dk}
\alignauthor Kåre Mølbak\\
       \affaddr{Statens Serum Institut}\\
       \email{KRM@ssi.dk}
\and  
\alignauthor Ingemar J. Cox\\
       \affaddr{University of Copenhagen}\\
       \email{ingemar.cox@di.ku.dk}
\alignauthor Christina Lioma\\
        \affaddr{University of Copenhagen}\\
		\email{c.lioma@di.ku.dk}
}

\maketitle
\begin{abstract}
Estimating vaccination uptake is an integral part of ensuring public health. 
It was recently shown that vaccination uptake can be estimated automatically from web data, instead of slowly collected clinical records or population surveys \cite{DalumHansen:2016:ELV:2983323.2983882}. All prior work in this area assumes that features of vaccination uptake  collected from the web are temporally regular. We present the first ever method to remove this assumption from vaccination uptake estimation: our method dynamically adapts to temporal fluctuations in time series web data used to estimate vaccination uptake. We show our method to outperform the state of the art compared to competitive baselines that use not only web data but also curated clinical data. This performance improvement is more pronounced for vaccines whose uptake has been irregular due to negative media attention (HPV-1 and HPV-2), problems in vaccine supply (DiTeKiPol), and targeted at children of 12 years old (whose vaccination is more irregular compared to younger children).


\end{abstract}

%
%

%
%

%
%



\section{Introduction and related work}

Vaccination programs are an efficient and cost effective method to improve public health. 
With sufficiently many people vaccinated the population gains herd immunity, meaning the disease cannot spread. Timely actions to avoid drops in vaccination coverage are therefore of great importance. Many countries have no registries of timely vaccination uptake information, but rely for example on yearly surveys. In such countries estimations of near real-time vaccination uptake based solely on web data are valuable. We extend prior work in this area \cite{DalumHansen:2016:ELV:2983323.2983882}, which showed that vaccination uptake can be estimated sufficiently accurately from web search data. 
Our extension consists of a new estimation method that adapts dynamically to temporal fluctuations in the signal (web search queries in our case) instead of assuming temporal stationarity as in \cite{DalumHansen:2016:ELV:2983323.2983882}. This contribution is novel within vaccination uptake estimation.


Linear models have been used previously to estimate health events, for instance by  combining data from multiple sources with an ensemble of decision trees \cite{santillana2015combining}, or, closer to our work, by using query frequencies for influenza like illness \cite{yang2015accurate} or vaccination uptake estimation \cite{DalumHansen:2016:ELV:2983323.2983882}. These approaches are designed for stationary time-series analysis, i.e. they assume data is generated by a stationary stochastic process. Our motivation is that vaccination uptake often does not follow stationary seasonal patterns. External events such as disease outbreaks, suspicion of adverse effects, or temporary vaccine shortages can alter uptake patterns for shorter or longer periods of time. Hence, while historical data is a good estimator in stable periods, as shown in \cite{DalumHansen:2016:ELV:2983323.2983882}, we reason that adapting the estimation to any unstability can reduce estimation error. We experimentally confirm this on all official children vaccines data used in Denmark between 2011 - 2016.  
%




\section{Aggregation with regression\\ trees for time series adaptation}
To account for seasonal non-stationarity, we use an online learning method, Aggregation Algorithm (AA) \cite{vovk2001competitive}, designed to automatically reduce estimation error in a changing environment. AA was recently used in time series prediction combined with an ensemble of ARIMA models \cite{jamil2016aggregation}. However, we reason that ARIMA models, or other traditional time series models, are not likely to be sufficient for vaccination uptake estimation in cases where: (i) there is more than one data source, e.g. vaccine uptake data and search frequency data, and (ii) when the time series data to be estimated are assumed to be unavailable (near real-time). 
To address these challenges, we combine AA with regression trees, motivated by recent research showing that random forests outperform ARIMA models on avian influenza prediction \cite{kane2014comparison}. A random forest, i.e. an ensemble of decision trees, is well suited for our problem since it is easy to extend to multiple data sources. 

Our method works as follows: We initially generate a set of regression trees. For each time step the ensemble of regression trees is retrained based on the initial set of trees and a weighted sum is used to make the estimation. AA is used to continuously update the weights of each tree. 
Each regression tree is trained based on a set of features and training samples. For each tree a feature set is drawn with replacement from the complete feature set.
Training samples are selected based on time-relative indices, where index 0 corresponds to the current time step. The indices are uniformly drawn with replacement from the interval $[0:s]$, where $s$ is a window size. We use trees with different window sizes to account for stationarity and non-stationarity of the signal. 

Our adaptive vaccination estimation algorithm is shown in Algorithm \ref{alg1}, where $\eta$ is the learning rate, RT a set of $N$ regression trees, $i$ the amount of initial training data and $y$ the vaccination uptake.
{ \footnotesize
\begin{algorithm}
	\caption{Adaptive time series estimation}
	\label{alg1}
	\begin{algorithmic}[1]
	\REQUIRE RT, $\eta$, $i$
	\STATE $W \leftarrow$ list with weights, initialize to be uniform
	\STATE $X \leftarrow$ list with the first $i$ training samples
	\STATE $Y \leftarrow$ list with the first $i$ observations of $y$
	\STATE $\hat{Y} \leftarrow$ empty list of estimations
	\STATE $t \leftarrow$ current time step, starting at $i+1$
	\WHILE {True}
	\STATE $x_t \leftarrow$ receive new observation from data stream
	\FOR {$n=0$ \TO $N$}
	\STATE Train RT$[n]$ using $X$ and $Y$
	\STATE $\hat{Y}_{temp}[n] \leftarrow$ estimation of RT$[n]$ given $x_t$
	\ENDFOR 
	\STATE $\hat{Y}[t] \leftarrow$ $\sum_{n=0}^N W[n]\cdot\hat{Y}_{temp}[n]$
	\STATE $Y[t] \leftarrow $ observed $y$ at time $t$
	\FOR{$n=0$ \TO $N$}
	\STATE $W[n] \leftarrow W[n]\cdot\exp(-\eta\cdot(\hat{Y}_{temp}[n]-Y[t])^2))$ 
	\ENDFOR 
	\STATE $W \leftarrow$ normalize $W$
	\STATE $X[t] \leftarrow x_t$
	\STATE $t \leftarrow t + 1$
	\ENDWHILE
	\end{algorithmic}
\end{algorithm}
}

\section{Evaluation}
To facilitate direct comparison, we evaluate our method on the same data as \cite{DalumHansen:2016:ELV:2983323.2983882}: monthly vaccination uptake of all official children vaccines in Denmark from January 2011 - June 2016. Vaccination uptake is defined as the total number of people vaccinated in a month divided by the birth cohort for that month. To estimate vaccination uptake, we use frequencies of web search queries extracted from Google Trends. We use the exact same frequencies of single term queries provided by \cite{DalumHansen:2016:ELV:2983323.2983882}. 


We compare to two baselines: (1) Linear regression with lasso regularization where the hyper-parameter is found using three fold cross-validation on the training data; (2) Linear regression with elastic net regularization where the two hyper-parameters are also selected using three fold cross-validation. 
We also include for reference two upper bounds corresponding to the best score reported in \cite{DalumHansen:2016:ELV:2983323.2983882} when using (i) only web data, and (ii) web data combined with clinical data. These scores are not theoretical upper bounds, but just the best scores across all methods evaluated in \cite{DalumHansen:2016:ELV:2983323.2983882}. We treat them as performance upper bounds because they do not correspond to any individual method, but to the best score per vaccine across all methods reported in \cite{DalumHansen:2016:ELV:2983323.2983882}. Neither baselines or upper bounds account for time-series adaptation, i.e. they all assume data stationarity.


The initial number of training samples, $i$, is set to 24. All algorithms are evaluated in a leave-one-out fashion, where all data prior to the data point being estimated is used for training. For our algorithm (ATSE) a parameter search is performed by randomly sampling from the following intervals: Window size interval 1-46, number of features derived from vaccination data 0-45, number of features derived from web data 0-30, number of regression trees 500-10000, $\eta$ between 0.001-0.25.

\begin{table}
    \centering
    \scalebox{0.95}{
    \begin{tabular}{l|r|r|r|r|r}
    Vaccine&LASS & EN & ATSE& UBW \cite{DalumHansen:2016:ELV:2983323.2983882}  & UBWC \cite{DalumHansen:2016:ELV:2983323.2983882} \\
        HPV-1		 & 14.6 & 13.8    &\bf10.0*	&11.5&9.3\\
        HPV-2		& 15.9  & 16.1    &\bf10.1&15.4&8.7\\
        MMR-1		& 12.9 & 12.9  &\bf12.6*&16.5&14.9\\
        MMR-2(4)		&15.5& 14.7   & \bf14.2 &12.4&12.3\\
        MMR-2(12)	& 21.7  & 21.4    &\bf16.0*&20.8&16.5\\
        DiTeKiPol-1	& 16.2  & 16.2  &\bf10.8&8.0&4.6\\
        DiTeKiPol-2	& 14.1 & 14.2  &\bf12.4&9.9&7.1\\
        DiTeKiPol-3	& 10.8 & 11.1  &\bf10.0*&17.1&16.4\\
        DiTeKiPol-4	&\bf13.7* & 14.3*   &14.4*&15.4&14.4\\
        PCV-1		&\bf 7.5 & 7.8   &10.0&7.7&5.2\\
        PCV-2		&9.6* 	& \bf9.5 		 &10.0&9.6&6.4 \\
        PCV-3		&\bf9.4*		& 9.5*		 &10.1*&10.3&6.6\\
    \end{tabular}
    }
    \caption{Estimation error when estimating vaccination uptake from web search queries with our method (ATSE), Lasso (LASS), Elastic Net (EN), and the two performance upper bounds of \cite{DalumHansen:2016:ELV:2983323.2983882} with web search (UBW) and web search and clinical data (UBWC). Bold marks best (excluding upper bounds). Asterisk marks better or equal to any upper bound.}
	\label{tab:results}
\end{table}

 Table~\ref{tab:results} displays the root mean squared error (RMSE) between the estimated vaccination uptake and the real vaccination uptake for all methods. 
Our method yields the overall best performance compared to the baselines (it outperforms all baselines for 8 out of 12 vaccines). Our method also outperforms the upper bounds of \cite{DalumHansen:2016:ELV:2983323.2983882} (any of the two) for 6 vaccines. This supports our reasoning that adapting the estimation to temporal fluctuations is a better strategy than assuming data stationarity. Our method yields the strongest performance improvements for HPV-1, HPV-2, MMR-2(12) and DiTeKiPol. All of these vaccines have temporally irregular uptake patterns, as explained next.
HPV-1 and HPV-2 have been subject to a heavy media debate in Denmark, with a subsequent drop in vaccinations. MMR-2(12) denotes the second MMR vaccine targeted 12 year-olds. As children grow, parents are less likely to follow the recommended vaccination schedule and fluctuations correlated with measles outbreaks are observed, thus making the time series less stationary. Lastly, in recent years there have been problems obtaining a sufficient supply of certain DiTeKiPol vaccines in Denmark, which might have forced people to postpone the initial vaccination, hence introducing irregularities in the signal. For PCV vaccines there have been no noted irregularities in their uptake patterns, which explains the slight drop in performance by our method compared to the baselines.


\section{Conclusion}

We presented an automatic method for near real time estimation of health events using web search query data. Our method combines an Aggregation Algorithm (AA) to automatically reduce estimation error in changing environments with regression trees. We applied our method to estimate vaccination uptake in all official Danish children vaccines, following  \cite{DalumHansen:2016:ELV:2983323.2983882}, and showed that our approach overall outperformed strong baselines that assumed data to be temporally regular. Our method was particularly strong estimating uptake for vaccines with known irregularities in their usage, such as HPV-1, HPV-2, MMR-2(12) and DiTeKiPol.

This work confirms recent findings that vaccination uptake can be automatically estimated only from web data, and further extends this area by accounting for irregular uptake patterns. 

\medskip

\noindent {\scriptsize Funded by IBM and the Danish Agency for Science \& Higher Education.}


\end{document}